\documentclass[runningheads,a4paper]{llncs}

\usepackage{amssymb}
\usepackage{amsmath}
\usepackage{graphicx}
\usepackage{epstopdf}
\usepackage{colortbl}

\usepackage{url}
\urldef{\mailsa}\path|{a.naumchev,b.meyer,v.rivera}@innopolis.ru|
\newcommand{\keywords}[1]{\par\addvspace\baselineskip
\noindent\keywordname\enspace\ignorespaces#1}

\usepackage{listings}
\usepackage{eiffel}
\usepackage{framed}
\usepackage{fancyvrb}
\usepackage{zed-csp}
\usepackage{caption}
\usepackage{enumerate}

\begin{document}
\lstset{language=Eiffel, morekeywords={agent}}

\mainmatter  

\title{Unifying Requirements and Code: an Example}


%
%
\author{Alexandr Naumchev, Bertrand Meyer*, Victor Rivera}
%

\institute{Innopolis University, Software Engineering Laboratory,\\
Innopolis, Russia\\
*Also ETH Z{\"u}rich\\
\mailsa\\
\url{http://university.innopolis.ru/}}

%
%

\maketitle

\begin{abstract}
Requirements and code, in conventional software engineering wisdom, belong to entirely different worlds. Is it possible to unify these two worlds? A unified framework could help make software easier to change and reuse. To explore the feasibility of such an approach, the case study reported here takes a classic example from the requirements engineering literature and describes it using a programming language framework to express both domain and machine properties. The paper describes the solution, discusses its benefits and limitations, and assesses its scalability.
\end{abstract}

\section{Introduction}
\label{intro}
According to the  standard view in software engineering, the tasks of requirements, design and implementation require distinct techniques and produce different artifacts.

What if instead of focusing on the differences we recognized the fundamental unity of the software construction process through all its stages? The principle of “seamlessness” (see e.g. \cite{meyer1988object}) follows from this assumption that the commonalities are more fundamental than the differences, and that it pays to use the same set of concepts, notations and tools throughout the development, from the most general and user-oriented initial steps down to the most technical tasks.

A consequence of the seamlessness principle is that requirements are just another software artifact, susceptible to many of the same techniques as code and design. In particular, assuming a modern programming language with powerful abstraction facilities, the requirements can be written in the same notation as the program.

The notion of multirequirements \cite{Meyer13Multi} adds to this principle the idea of using several interleaved descriptions: natural language, graphical, and formal (Eiffel text) serving as the reference.

How realistic is the seamless multirequirements approach, what are its limits, and what benefits does it bring? To help answer this question, the present article takes the example used in a classic paper of the requirements literature, Jackson's and Zave's zoo control system, and describes it entirely in a seamless style, including the formal constraints that form a key part of the original article.

The goal of the paper is not advocacy but experimentation. The advocacy is present in the earlier references cited above. We practice a seamless approach to software construction and consider it fruitful, but the present discussion does not attempt to establish its superiority; rather it starts from the seamlessness hypothesis - in particular, the hypothesis that a single notation, Eiffel, is applicable to requirements analysis just as much as to programming - and applies this hypothesis fully and consistently to a significant example. While we draw some conclusions, the important part is the result of the experiment as presented here, enabling readers to form their own conclusions as to the benefits and limits of the approach.

Section \ref{problems} briefly explains why it is interesting to put into question the traditional separation between software development tasks. Section \ref{approach} proposes an approach to unify software development tasks by combining the approaches described in \cite{Meyer13Multi} and \cite{jackson1995deriving}. Section \ref{background} introduces some theoretical and technical background. Section \ref{example} presents the approach applied to an example. Finally, Section \ref{conclusion} concludes and mentions future work.

\subsection{Summary of Contributions}
Experimentation mentioned at the end of Section \ref{intro} resulted in the following key outcomes.
\begin{itemize}
\item An evidence suggesting that it is possible to use Multirequirements approach \cite{Meyer13Multi} for describing cyber-physical systems like zoo turnstile controller. At the same time, different types of exemplar statements goes far beyond just the relational statements used in \cite{Meyer13Multi}.
\item An evidence suggesting that a real programming language notation may be even more expressive than most of the popular formal notations. Section \ref{opt2} contains all the details.
\item An example showing how object orientation helps to effectively manage complexity in specifications. The approach used in \cite{jackson1995deriving}, where the specification is basically a linear list of statements, does not scale to the case of large systems, when the number of requirements is too big. Object orientation provides a way to relate the conceptual objects so that the resulting specification will be scattered across the classes in an intuitive way.
\end{itemize}

\section{The drawbacks of too much separation of concerns}
\label{problems}
Historically, there was a reason for emphasizing the distinction between development tasks. The goal was to highlight the specific needs of requirements and design, moving away from the ``code first, think later" way of building software. But as the precepts of software engineering have gained wide acceptance and programming languages have moved from low-level machine-coding notations to descriptive formalisms with high expressive power, the reverse approach is worth exploring: instead of emphasizing the differences, show the fundamental unity of the software process.

The traditional approach is subject to five criticisms.

\begin{enumerate}[i)]
\item \label{natural_text_problems} Insufficient information. Requirements analysts do not know what details are important for developers. They are good at expressing customer needs in a form the customer is ready to sign, but they typically do not know what is implementable and what is not. \cite{meyer1985formalism} discusses some typical flaws of natural language requirements specifications.
\item \label{communication_problem} Lack of communication. When developers see ambiguous or contradictory elements in the requirements, they will not always go back and ask, but will often interpret the requirement according to their own understanding, which may or may not coincide with user wishes.
\item \label{conversation} Impedance mismatches \cite{meyer1988object}. The use of different formalisms at different stages requires translations and creates risks of mistakes.
\item Impediment to change. With different formalisms, it is difficult \cite{meyer1988object} to ensure that a change at one level is reflected at other levels.
\item Impediment to reuse. The presence of requirements as a document specific to each project may mask the commonality between projects and make the team miss potential reuse of existing developments.
\end{enumerate}

\section{A seamless approach}
\label{approach}
\subsection{Unifying processes}
Consideration of the problems listed above leads to trying a completely different approach, which recognizes that beyond the obvious differences between tasks of software  development  they share fundamental needs, concepts, principles, techniques. In particular, they can be addressed through a common notation.  Modern programming languages are not just coding tools to talk to a machine, but powerful tools  for expressing abstract concepts and modeling complex systems. The Eiffel notation used in the present work uses object-oriented  principles of classes, genericity, polymorphism and inheritance, which have proved adept at describing  sophisticated systems (independently of their technical programming aspects) in a modular, flexible, reusable and evolutionary way. Thanks to the presence of Design by Contract mechanisms, it can describe not only the structure of systems but their abstract semantics.
\subsection{The Hypothesis}
\label{hypothesis}
The hypothesis explored in this paper, in light of the above analysis, is that it is possible to design a software development process that:
\begin{enumerate}[i)]
\item \label{re_design} Uses for requirements the same notation and tools as for design and implementation.
\item \label{impl} Links the resulting documents (requirements, design, code) together, ensuring a major goal of software engineering: traceability.
\item \label{proof} Makes it possible to prove, formally, the correctness of the implementation against the specification.
\item \label{continuity} Supports extendibility  by ensuring that small changes in the requirements will cause a proportionally small change in the design and the implementation.
\end{enumerate}

\subsection{How to Test the Hypothesis}
The present work relies on the following scenario for testing the preceding hypothesis at least in part:
\begin{enumerate}[i)]
\item \label{propose} Propose a candidate process.
\item \label{projects} Select examples and apply the process.
\item \label{test} Analyze the outcome.
\end{enumerate}
\cite{Meyer13Multi} sketches such a process, based on using object orientation for representing the relationships between the conceptual objects in the requirements document. The basic idea was to have an object-oriented code along with the natural language description of a requirement. It is also possible to represent each code fragment graphically as a BON diagram \cite{walden1995seamless}.

\cite{Meyer13Multi}, however, uses as example the very notion of requirements process. In other words, it is self-referential. This confers (we hope) a certain elegance to the example, but makes it look artificial. In the present paper we take a more standard example, coming from a classic requirements paper by Jackson and Zave \cite{jackson1995deriving}.

More precisely, the requirements from the example are represented using the model-based \cite{polikarpova2014specified} contracts-equipped \cite{meyer2009touch} object-oriented \cite{meyer1988object} notation (Eiffel).

\section{Theoretical and Technical Background}
\label{background}
\subsection{Design By Contract}
Work \cite{meyer2009touch} gives a comprehensive description of Design By Contract. Design By Contract integrates Hoare-style assertions \cite{hoare1969axiomatic} within object-oriented programs \cite{meyer1988object} constraining the data that run time objects hold. This approach equips each class feature (member) with a predicate expression, that specify its behavior, in the form of pre- and postcondition. The postcondition has to hold whenever the precondition held and the feature finished its computation before the program execution process invokes the next feature. Design By Contract equips the class itself with an invariant predicate expression which holds in all states of the corresponding objects.
\subsection{Model-Based Contracts}
\label{model_based_contracts}
If classical contracts are for constraining the data that run time objects actually hold, model-based contracts are ``meta" contracts for constraining the objects as mathematical entities (sets, sequences, bags, relations etc.), and an execution process does not instantiate the corresponding mathematical representations at run time as parts of the objects.  Model-Based Contracts are useful when it is not possible to capture all the nuances by means of classical contracts. The PhD thesis \cite{polikarpova2014specified} gives some examples of such situations and a comprehensive description of the concept.
\subsection{AutoProof}
The AutoProof \cite{tschannen2012automatic} tool is capable of formally proving the correctness of contract-equipped object-oriented programs, both classical and model-based. AutoProof proves for every routine that the conjunction of the precondition and the class invariant before invocation ensures the conjunction of the postcondition and the class invariant after invocation. The class is verified if and only if all the class features are verified.

\section{Unifying the Two Worlds: an Example}
\label{example}
Avoiding the problems analyzed in Section \ref{problems} means unifying the worlds of requirements and code in a unified framework. This section illustrates the approach. It takes the example from the work \cite{jackson1995deriving} and shows how to express requirements of various types in the style of work \cite{Meyer13Multi} - namely, using Eiffel as a formal specification language for expressing each requirement. Originally the authors used this example to demonstrate the process of deriving specifications from requirements, and the unified approach captures all the  nuances of this process.
\subsection{Example Overview}
\begin{figure}[!t]
\begin{tabular}{m{0.55\textwidth} m{0.52\textwidth}}
\begin{lstlisting}[basicstyle=\scriptsize]
class ZOO
feature
  turnstile: TURNSTILE
end

class TURNSTILE
feature
  coinslot: COINSLOT
  barrier: BARRIER
invariant
  coinslot.turnstile = Current
  barrier.turnstile = Current
end
\end{lstlisting}
&
\begin{lstlisting}[basicstyle=\scriptsize]
class COINSLOT
feature
  turnstile: TURNSTILE
invariant
  turnstile.coinslot = Current
end

class BARRIER
feature
  turnstile: TURNSTILE
invariant
  turnstile.barrier = Current
end
\end{lstlisting}
\end{tabular}
\caption{Expressing the context formally}
\label{fig:context}
\end{figure}
The authors of \cite{jackson1995deriving} start with giving the overall context: \textit{``...Our small example concerns the control of a turnstile at the entry to a zoo. The turnstile consists of a rotating barrier and a coin slot, and is fitted with an electrical interface..."}
This small paragraph mostly describes the relationships between the conceptual objects. Figure \ref{fig:context} contains specification of the context in the style of work \cite{Meyer13Multi}.

Translating the specification from Figure \ref{fig:context} back to natural language using the object-oriented semantics results in almost the same initial description: ``A ZOO has a TURNSTILE turnstile; a TURNSTILE has a COINSLOT coinslot and a BARRIER barrier so that coinslot has Current TURNSTILE as turnstile and barrier has Current TURNSTILE as turnstile..." COINSLOT and BARRIER hold references to the TURNSTILE instances in order to capture the \textit{``electrical interface"} phenomena: the word ``interface" means something over which the parties are able to communicate with each other; communicating means sending messages to each other, and to send message to someone in the object-oriented world is to take a reference to the object and perform a qualified call on it. So at the very least the parties should hold references to each other to be able to communicate in two directions.
\subsection{The Designation Set}
\begin{figure}
\begin{itemize}
\item \textbf{Push}($e$): In event $e$ a visitor pushes the \textbf{barrier} to its intermediate position
\item \textbf{Enter}($e$): In event $e$ a visitor pushes the barrier fully home and so gains entry to the \textbf{zoo}
\item \textbf{Coin}($e$): In event $e$ a valid coin is inserted into the \textbf{coin slot}
\item \textbf{Lock}($e$): In event $e$ the \textbf{turnstile} receives a locking signal
\item \textbf{Unlock}($e$): In event $e$ the \textbf{turnstile} receives an unlocking signal
\end{itemize}
\caption{The Zoo Turnstile example designation set}
\label{fig:zoo_design_set}
\end{figure}
After stating the problem context the authors of \cite{jackson1995deriving} describe the \textit{designation set}. Each designation basically corresponds to a separate type of events observed in the problem area. The authors give the designations as a set of predicates as in Figure \ref{fig:zoo_design_set}.
\begin{figure}[!t]
\begin{tabular}{m{0.55\textwidth} m{0.52\textwidth}}
\begin{lstlisting}[basicstyle=\scriptsize]
note
  model: enters
deferred class ZOO
feature
  enter
  deferred
  ensure
    enters.but_last ~ old enters
    enters.last > old enters.last
  end
  enters: MML_SEQUENCE[INTEGER_64]
end

note
  model: locks, unlocks
deferred class TURNSTILE
feature
  lock
  deferred
  ensure
    locks.but_last ~ old locks
    locks.last > old locks.last
  end
  unlock
  deferred
  ensure
    unlocks.but_last ~ old unlocks
    unlocks.last > old unlocks.last
  end
  locks: MML_SEQUENCE[INTEGER_64]
  unlocks: MML_SEQUENCE[INTEGER_64]
end
\end{lstlisting}
&
\begin{lstlisting}[basicstyle=\scriptsize]
note
  model: coins
deferred class COINSLOT
feature
  coin
  deferred
  ensure
    coins.but_last ~ old coins
    coins.last > old coins.last
  end
  coins: MML_SEQUENCE[INTEGER_64]
end

note
  model: pushes
deferred class BARRIER
feature
  push
  deferred
  ensure
    pushes.but_last ~ old pushes
    pushes.last > old pushes.last
  end
  pushes: MML_SEQUENCE[INTEGER_64]
end
\end{lstlisting}
\end{tabular}
\caption{Specifying the designation set formally}
\label{fig:design_set}
\end{figure}
Figure \ref{fig:design_set} is an Eiffel implementation of each designation set described in Figure \ref{fig:zoo_design_set}. The implementation uses Eiffel features names as labels for the events types. The natural language descriptions from Figure \ref{fig:zoo_design_set} provide heuristics on which feature should be added to which class (Figure \ref{fig:zoo_design_set} highlights the correspondence with \textbf{bold}). Each event type has an associated history - a sequence of moments in time when the events of this particular type occurred. For example, $enters: MML\_SEQUENCE[INTEGER\_64]$ (in Figure \ref{fig:design_set}) is a sequence of moments in time expressed in milliseconds when events of type $enter$ took place. $MML\_SEQUENCE$ is a class from the $MML$ (Mathematical Modeling Library) and denotes mathematical sequence. $MML$ contains special classes for expressing model-based contracts. Although it is possible to instantiate some simple objects from these classes (like a sequence containing one element), the instances will not be modifiable. The $model$ annotation is the Eiffel mechanism to represent model-based contracts (introduced in section \ref{model_based_contracts}). For instance, expression $model: enters$ in Figure  \ref{fig:design_set} gives a hint that $enters$ feature will be used for expressing the model-based part of the contract.

The $deferred$ keyword states that the specification gives only formal definitions of the events (in terms of pre- and postconditions \cite{hoare1969axiomatic}) and does not give the corresponding operational reactions of the machine on the events. The $ensure$ clause is the postcondition of the feature. It describes how the system changes after reacting on an event of the corresponding type. These specifications are intuitively plausible: an event occurrence should result in extending the corresponding history with the moment in time when the event took place, and the time of the new event should be strictly bigger than the time of the previous event, as shown, for instance, by the postcondition in feature $unlock$ of Figure \ref{fig:design_set}. The keyword {\bf old} is used to indicate expressions that must be evaluated in the pre-state of the routine, and $\char126$ makes a comparison by value.
\subsection{Shared Phenomena}
The authors of \cite{jackson1995deriving} introduce the notion of shared phenomena - that is, the phenomena visible to both the world (the environment) and the machine (the notions of the world and the machine were introduced by Jackson in \cite{jackson1995world}). In the present approach this notion is covered by using the ``has a" relationships between the $ZOO$ and the $TURNSTILE$ classes, accompanied with the model-based contracts. Namely, since a $ZOO$ has a turnstile as its feature, it can see any phenomena hosted by the turnstile: $locks, unlocks, coins, pushes$; since a $TURNSTILE$ does not hold any references to a $ZOO$, it can not observe nor control the $enter$ events modeled by $ZOO$.
\subsection{Specifying the System}
Work \cite{jackson1995deriving} introduces a set of criteria by means of which it is possible to identify whether the machine is specified or not. One of the criteria states that all requirements should be expressed in terms of shared phenomena only. Requirements refinement is the process of converting the requirements stated in terms of both shared and non-shared phenomena to the form in which they are expressed in terms of shared phenomena only. Refinement process consists of identifying some laws, which hold in the environment regardless of the machine behaviour, and constraining the machine behaviour. The resulting constraints imposed on the machine together with the laws of the environment should logically imply the requirements stated in the beginning.

The authors of \cite{jackson1995deriving} state that the laws of the environment are always expressed in the indicative mood, while the restrictions imposed on the machine behavior are expressed in the optative mood.

All properties of the problem derived in \cite{jackson1995deriving} - be they optative or indicative descriptions - can be conceptually divided into the two main categories.
\subsubsection{Properties which hold at any moment in time:}
\begin{figure}[!t]
\begin{tabular}{m{0.55\textwidth} m{0.52\textwidth}}
\begin{lstlisting}[basicstyle=\scriptsize]
deferred class ZOO
feature
  turnstile: TURNSTILE
  enters: MML_SEQUENCE[INTEGER_64]
invariant
  enters.count <= turnstile.coinslot.coins.count
end
\end{lstlisting}
\caption{Entries should never exceed payments}
\label{fig:opt1}
\end{tabular}
\end{figure}
an example of such property is the $OPT1$ requirement (expressed in Figure \ref{fig:opt1}) saying that entries should never exceed payments (the authors of \cite{jackson1995deriving} use $OPT*$ for labeling properties expressed in an optative mood). Within the present approach this requirement can be expressed in the following way. The ``something always holds" semantics fits perfectly into the semantics of Eiffel invariant: ``something holds in all states of the object", as expressed in Figure \ref{fig:opt1}.

\subsubsection{Properties which hold depending on the type of the next event to occur:}
\begin{figure}[!t]
\begin{tabular}{m{0.55\textwidth} m{0.52\textwidth}}
\begin{lstlisting}[basicstyle=\scriptsize]
deferred class BARRIER
feature
  push
  require
    not turnstile.unlocks.is_empty
    (not turnstile.locks.is_empty) implies (turnstile.unlocks.last >
                                           turnstile.locks.last)
  deferred
  end
end
\end{lstlisting}
\caption{It is impossible to use locked turnstile}
\label{fig:ind2}
\end{tabular}
\end{figure}
the indicative property $IND2$ saying that it is impossible to push the barrier if the turnstile is locked will serve as an example (the authors of \cite{jackson1995deriving} use $IND*$ for labeling properties expressed in the indicative mood). Figure \ref{fig:ind2} depicts the corresponding specification.
The initial description is divided into the two different claims: first, the turnstile should be unlocked at least once, and second, if the turnstile has ever been locked, the last unlock should have occurred later than the last lock.
\subsubsection{Real Time Properties:}
\begin{figure}[!t]
\begin{tabular}{m{0.55\textwidth} m{0.52\textwidth}}
\begin{lstlisting}[basicstyle=\scriptsize]
deferred class BARRIER
feature
  turnstile: TURNSTILE
  push
  deferred
  ensure
    ((old turnstile.unlocks.last > old turnstile.locks.last) and
     (pushes.count = turnstile.coinslot.coins.count))
      implies (turnstile.locks.last > pushes.last and
               (turnstile.locks.last - pushes.last) < 760)
  end
  pushes: MML_SEQUENCE[INTEGER_64]
end
\end{lstlisting}
\caption{The machine locks the turnstile timely}
\label{fig:opt7}
\end{tabular}
\end{figure}
the authors of \cite{jackson1995deriving} derive several timing constraints on the events processing. For example, the $OPT7$ requirement says that the amount of time between the moment when the number of the barrier pushes becomes equal to the number of coins inserted and the moment when the machine locks the turnstile should be less than 760 milliseconds. This is basically a constraint for the reaction on the $push$ event: if the next $push$ event uses the last coin, the machine should ensure that the turnstile is locked in a timely fashion, so that a human being will not have time to enter without paying. The $760$ quantity reflects the fact that it takes at least 760 milliseconds for a human being to rotate the barrier completely and enter the Zoo.

Taking this reasoning into consideration, the present specification approach handles the timing constraint by putting it into the $push$ feature postcondition (as depicted in Figure \ref{fig:opt7}). The antecedent of the implication assumes the situation when before the $push$ event the turnstile was locked ($old turnstile.unlocks.last > old turnstile.locks.last$ expression in Figure \ref{fig:opt7}), and after the event occurrence the number of barrier pushes became equal to the number of coins inserted ($pushes.count = turnstile.coinslot.coins.count$ expression in Figure \ref{fig:opt7}). The consequent reflects the requirement that, having in place the situation that the antecedent describes, there should be a $lock$ event which is more late than the last $push$ event ($turnstile.locks.last > pushes.last$ expression in Figure \ref{fig:opt7}), and the distance between them should be less than 760 milliseconds ($(turnstile.locks.last - pushes.last) < 760$ expression in Figure \ref{fig:opt7}).

\subsection{Specifying the ``Unspecifiable"}
\label{opt2}
\begin{figure}[!t]
\begin{tabular}{m{0.55\textwidth} m{0.52\textwidth}}
\begin{lstlisting}[basicstyle=\scriptsize]
deferred class ZOO
feature
  turnstile: TURNSTILE_ABSTRACT
  enter
  deferred
  end
  enters: MML_SEQUENCE[INTEGER_64]
invariant
  turnstile.coinslot.coins.count > enters.count implies
    (agent enter).precondition
end
\end{lstlisting}
\caption{The turnstile let people who pay enter}
\label{fig:opt2}
\end{tabular}
\end{figure}
One of the requirements mentioned in \cite{jackson1995deriving} was $OPT2$ saying that the visitors who pay are not prevented from entering the Zoo. The authors give only informal statement of this requirement:
$\forall v,m,n @ ((Enter\#(v,m) \land Coin\#(v,n) \land (m < n)) \implies 'The\ machine\ will\ not\ prevent\ another\ Enter\ event'$.

The antecedent of this implication should be read like ``the number of entries is less than the number of coins inserted". The authors of \cite{jackson1995deriving} do not formalize the consequent and leave it in the natural language form. The present specification approach handles this requirement using standard Eiffel mechanism called \textit{agents} (see Figure \ref{fig:opt2}).

The $agent$ clause treats a feature (the $enter$ feature in this particular case) as a separate object so that the feature precondition becomes one of the boolean-type features of the resulting object.
\section{Conclusion}
\label{conclusion}

Software construction involves different activities. Typically these activities are performed separately. For instance, requirements and code, as  developed nowadays, seem to belong to different  worlds. The case study reported in this paper shows the feasibility of unifying requirements and code in a single framework.

This paper takes the classic Zoo Turnstile example \cite{jackson1995deriving} and implements it using Eiffel programming language. Eiffel is used not just to express the domain properties but also the properties of the machine \cite{jackson1995world}, enabling users to combine requirements and code in a single framework. This paper does not present the complete implementation of the example due to limited space. Full implementation can be reached in the GitHub project \cite{anaumchev2015zoo}.

The specification approach presented in this work is suitable not only for formalizing the statements that \cite{jackson1995deriving} formalizes, but also for formalizing those which are not possible to formalize with classical instruments like predicate or temporal logic (like $OPT2$ requirement, see Figure \ref{fig:opt2}).

The present approach is not only expressively powerful - it enables smooth transition to design and implementation. GitHub project \cite{anaumchev2015zoo} contains a continuation of the present work in the form of a complete implementation of the Zoo Turnstile example.

In order to understand the benefits of the present approach better it seems feasible to evaluate it against the hypothesis stated in Section \ref{hypothesis}:
\begin{enumerate}[i)]
\item \label{simultaneity} Unity of software development tasks: indeed, all the code fragments corresponding to different specification items merged together will bring a complete design solution available at \cite{anaumchev2015zoo} (the classes ending with ``\_abstract").
\item Traceability between the specification and the implementation: the classes ending with ``\_concrete" available at \cite{anaumchev2015zoo} contain the implementation and relate to the specification classes by means of inheritance.
\item Provability of the classes: the AutoProof system \cite{tschannen2012automatic} is capable of formally proving both classical and model-based contracts in Eiffel. However, it is not yet capable of proving "higher-level" agents-based contracts like the one used in Figure \ref{fig:opt2} for expressing requirement $OPT2$ from the work \cite{jackson1995deriving}. Adding this functionality to AutoProof is one of the next work items.
\item Extendibility of the solution: since Eiffel artifacts used in the formalizations of the requirements items correspond to their natural language counterparts directly, it is visible right away how a change in one representation will affect the second.
\end{enumerate}

Speaking about scalability of the approach, a formal representation of a requirements item specified with Eiffel is as big as the scope of the item and its natural language description are, so the overall complexity of the final document should not depend on the size of the project. Anyway, this is something to test by applying the approach to a bigger project.
\subsection{Future Work}
The future actions plan include:
\begin{enumerate}[i)]
\item \label{consistency} to prove formally that the specifications are consistent. In particular to ensure that the features specifications preserve the invariants of their home classes; to ensure that the invariants are self-consistent. For example it should not be possible for $P(x)$ and $\neg P(x)$ to hold at the same time.
\item \label{bon} to extend the BON notation \cite{walden1995seamless} so that it will be capable of expressing model-based contracts.
\item to design machinery for translating model-based contract-oriented requirements to their natural language counterpart so that the result will be recognizable by a human being.
\item to apply the approach to a bigger project.
\item \label{fix_autoproof} to extend AutoProof technology \cite{tschannen2012automatic} so that it will be able to handle agents in specifications (like in Figure \ref{fig:opt2}).
\end{enumerate}
It seems feasible to utilize AutoProof technology \cite{tschannen2012automatic} for achieving goal (\ref{consistency}). AutoProof is already capable of proving that a feature implementation preserves its specification (except specifications with agents), and it seems logical to empower it with the capabilities for working solely on the specifications level. Work \cite{nordio2009proofs} contains a formal proof that it is possible to achieve goal (\ref{fix_autoproof}).

As a result of implementing the plan a powerful framework for expressing all possible views on the software under construction should emerge. The threshold of success includes the possibility to generate the specification classes (their names end with ``\_abstract") available at \cite{anaumchev2015zoo} automatically, using requirements documents produced according to the present process as input.


\subsubsection*{Acknowledgment}
This work has been supported by the Russian Ministry of education and science with the project "Development of new generation of cloudy technologies of storage and data control with the integrated security system and the guaranteed level of access and fault tolerance" (agreement: 14.612.21.0001, ID: RFMEFI61214X0001). Also, the authors would like to thank their colleagues Alexander Chichigin and Dr. Manuel Mazzara from the Innopolis University Software Engineering Laboratory for their invaluable feedback.

{{{
\bibliographystyle {ieeetr}
\bibliography {psi}
}}}

\end{document}